\documentclass[longauth]{aa} 
\usepackage{euclid}
\usepackage[colorlinks=true,linkcolor=blue,citecolor=blue]{hyperref}
\usepackage[dvipsnames]{xcolor}
\usepackage{graphicx}
\graphicspath{{graphics}}
\usepackage{txfonts}
\usepackage[separate-uncertainty=true, list-units=repeat]{siunitx}
\usepackage{subcaption}
\usepackage{hyperref}
\usepackage[inline]{enumitem}
\usepackage{physics} 	
\usepackage{multirow}
\usepackage{bm}
\usepackage{soul}
\usepackage[flushleft]{threeparttable}
\usepackage{pdflscape}
\usepackage{graphicx}

\newcommand{\LePHARE}{\textsc{LePHARE}}
\newcommand{\Cigale}{\textsc{Cigale}}

\DeclareSIUnit{\Msun}{M_\odot}
\DeclareSIUnit{\year}{yr}
\DeclareSIUnit{\pc}{pc}
\DeclareSIUnit{\mag}{mag}
\DeclareSIUnit{\mas}{mas}
\DeclareSIUnit{\dex}{dex}
\DeclareSIUnit{\jansky}{Jy}
\DeclareSIUnit{\kpc}{kpc}

\begin{document} 

   \title{The COSMOS-Web ring: Spectroscopic confirmation of the background source at $z=5.1$}
   \subtitle{}  

\author{
Marko Shuntov\inst{\ref{DAWN},\ref{NBI}}\thanks{Corresponding authors: marko.shuntov@nbi.ku.dk, shuji@dtu.dk} \and
Shuowen Jin\inst{\ref{DAWN},\ref{DTU}}\footnotemark[1] \and
Wilfried~Mercier \inst{\ref{LAM}} \and%
Jeyhan S.~Kartaltepe\inst{\ref{Rochester}} \and
Rebecca Larson\inst{\ref{Rochester}} \and%
Ali Ahmad Khostovan\inst{\ref{Kentucky},\ref{Rochester}}
Raphaël Gavazzi\inst{\ref{LAM},\ref{IAP}} \and
James W.~Nightingale\inst{\ref{DurhamComputation}} \and%
Olivier Ilbert\inst{\ref{LAM}} \and
Rafael Arango-Toro\inst{\ref{LAM}} \and
Maximilien Franco\inst{\ref{IAP}} \and
Hollis B.~Akins\inst{\ref{UAT}} \and
Caitlin M.~Casey\inst{\ref{ucsb}, \ref{UAT}, \ref{DAWN}} \and
Henry Joy McCracken\inst{\ref{IAP}} \and
Laure Ciesla\inst{\ref{LAM}} \and
Georgios E.~Magdis\inst{\ref{DAWN},\ref{DTU}} \and
Aristeidis Amvrosiadis\inst{\ref{DurhamComputation}} \and%
Andrea Enia\inst{\ref{Bologna}, \ref{INAF}} \and%
Andreas L.~Faisst\inst{\ref{Caltech}} \and
Anton M.~Koekemoer\inst{\ref{STScI}} \and
Clotilde Laigle\inst{\ref{IAP}} \and%
Damien Le~Borgne\inst{\ref{IAP}} \and%
Richard Massey\inst{\ref{DurhamComputation}, \ref{DurhamExtragal}} \and%
Thibaud Moutard\inst{\ref{LAM}}\and%
Mattia Vaccari\inst{\ref{UCT}}
}

\institute{  
   Cosmic Dawn Center (DAWN), Denmark \label{DAWN}  
   \and
   Niels Bohr Institute, University of Copenhagen, Jagtvej 128, 2200 Copenhagen, Denmark \label{NBI}%
   \and
   DTU Space, Technical University of Denmark, Elektrovej 327, 2800 Kgs. Lyngby, Denmark \label{DTU}
   \and
   Aix Marseille Univ, CNRS, CNES, LAM, Marseille, France \label{LAM}  
   \and
   Laboratory for Multiwavelength Astrophysics, School of Physics and Astronomy, Rochester Institute of Technology, 84 Lomb Memorial Drive, Rochester, NY 14623, USA \label{Rochester}  
   \and
   Department of Physics and Astronomy, University of Kentucky, 505 Rose Street, Lexington, KY 40506, USA \label{Kentucky}
   \and
   Institut d’Astrophysique de Paris, UMR 7095, CNRS, Sorbonne Université, 98 bis boulevard Arago, F-75014 Paris, France \label{IAP}  
   \and
   Institute for Computational Cosmology, Durham University, South Road, Durham DH1 3LE, UK\label{DurhamComputation}%
   \and%
   The University of Texas at Austin, 2515 Speedway Blvd Stop C1400, Austin, TX 78712, USA\label{UAT}  
   \and   
   Broida Hall, Department of Physics, University of California, Santa Barbara, Santa Barbara, CA, 93106 USA\label{ucsb}%
   \and
   University of Bologna - Department of Physics and Astronomy “Augusto Righi” (DIFA), Via Gobetti 93/2, I-40129, Bologna, Italy\label{Bologna}%
   \and%
   INAF - Osservatorio di Astrofisica e Scienza dello Spazio, Via Gobetti 93/3, I-40129, Bologna, Italy\label{INAF}
   \and
   Caltech/IPAC, 1200 E. California Blvd. Pasadena, CA 91125, USA\label{Caltech}%
   \and%
   Space Telescope Science Institute, 3700 San Martin Drive, Baltimore, MD 21218, USA\label{STScI}%
   \and%
   Centre for Extragalactic Astronomy, Durham University, South Road, Durham DH1 3LE, UK\label{DurhamExtragal}%
   \and%
   University of Cape Town\label{UCT}%
   }

   \date{Received ; accepted }

 
  \abstract
  {
  We report the spectroscopic confirmation of the background source of the most distant Einstein ring known to date, the COSMOS-Web ring. This system consists of a complete Einstein ring at $z=5.1$, lensed by a massive early-type galaxy at $z\sim2$. 
  The redshift $z=5.1043\pm0.0004$ is unambiguously identified with our NOEMA and Keck/MOSFIRE spectroscopy, where the NOEMA observations reveal the CO(4-3) and CO(5-4) lines at $>8\,\sigma$, and the MOSFIRE data detect [O\textsc{ii}] at $\sim 6\,\sigma$.
  Using multi-wavelength photometry spanning near-infrared to radio bands, we find that the lensed galaxy is a dust-obscured starburst ($M_{\star} \sim 1.8\times10^{10}\,\si{\Msun}$, ${\rm SFR_{IR}\sim 60\,\si{\Msun} ~yr^{-1}}$) with high star-formation efficiency (gas depletion time $\tau_{\rm dep}<100~$Myr) as indicated by the [C\textsc{i}](1-0) non-detection. The redshift confirmation revalidates that the total lens mass budget within the Einstein radius is fully accounted for  by the stellar and dark matter components, without the need of modifying the initial mass function or dark matter distribution profile.
  This work paves the way for detailed studies and future follow-ups of this unique lensing system, providing an ideal laboratory for studying mass distribution at $z\sim2$ and physical conditions of star formation at $z\sim5$.
  }
   


   \maketitle

%

\section{Introduction}

Galaxy-galaxy lensing is a powerful tool for studying both the mass distribution around the foreground lens, and the physical conditions of the highly magnified background source.
In an extreme case of a perfect alignment, the lensing system produces an Einstein ring, a circular image of the background sources, which is ideal for spatially resolved studies of the interstellar medium (ISM) conditions of the lensed sources.
Throughout the last decade, serendipitous discoveries of strong lenses have been pushed towards increasingly higher redshifts. To date, the most distant spectroscopically confirmed lenses are at $z\sim1.6$ \citep{Wong14, Canameras17, Ciesla20}. Among these spectacular systems, strongly magnified dusty star-forming galaxies (DSFGs) have opened a window into dust-obscured and moderate star-formation activities at high redshifts \citep[][]{Vieira2013,Rizzo2020,Rizzo2021,Berta2021lens,Hamed2021,Cox2023zGal} and provided us with unprecedented deep insights into their ISM, kinematics, and dark matter properties \citep[e.g.][]{Canameras17, Yang2017, Cava2018, Ciesla20, Rizzo2020, Rizzo2021, Smail2023,Liu2024NatAs}.

Recently, \cite{Mercier2024Ring} reported the discovery of potentially the most distant lensing system from JWST imaging, the COSMOS-Web ring, in the JWST COSMOS-Web survey \citep[GO\#1727, PIs: Casey \& Kartaltepe;][]{Casey2023}. The system consists of a massive ($M_\star \sim \SI{e11}{\Msun}$) and quiescent (${\rm sSFR} \lesssim \SI{e-13}{\per\year}$) early-type galaxy (ETG) at $z_{\rm phot} \approx 2$. Around it is a complete Einstein ring which is formed by the deflection of light from a background source, potentially a DSFG at $z_{\rm phot} \approx 5$. 
The system was also discovered independently by \cite{vanDokkum2024Ring}. However, they derived a lower $z_{\rm phot}\sim3$ photometric redshift for the background source. The lower photometric redshift for the background source would imply a significantly higher total lensing mass than for a higher redshift solution. Thus, the exact redshift configuration of this lensing system, together with stellar mass and total dark matter mass estimates, potentially have important physical implications on the initial mass function (IMF), and even the nature of dark matter \citep[e.g.][]{Kong2024}. Therefore, it is critical to spectroscopically confirm the redshift of the background source. 

s Letter, we report the spectroscopic confirmation of the COSMOS-Web ring background source at $z=5.10$. We adopt a standard $\Lambda$CDM cosmology with $H_0=70$\,km\,s$^{-1}$\,Mpc$^{-1}$, $\Omega_{\rm m}=0.3$, and $\Omega_{\Lambda}=0.7$. Physical parameters are estimated assuming a \citet{Chabrier03} initial mass function (IMF). The magnitudes are expressed in the AB system \citep{Oke74}.

\section{Observations} \label{sec:observations}

\begin{figure*}[ht!]
\setlength{\abovecaptionskip}{-0.1cm}
  \centering
  \begin{subfigure}{0.56\textwidth}
    \includegraphics[width=\linewidth]{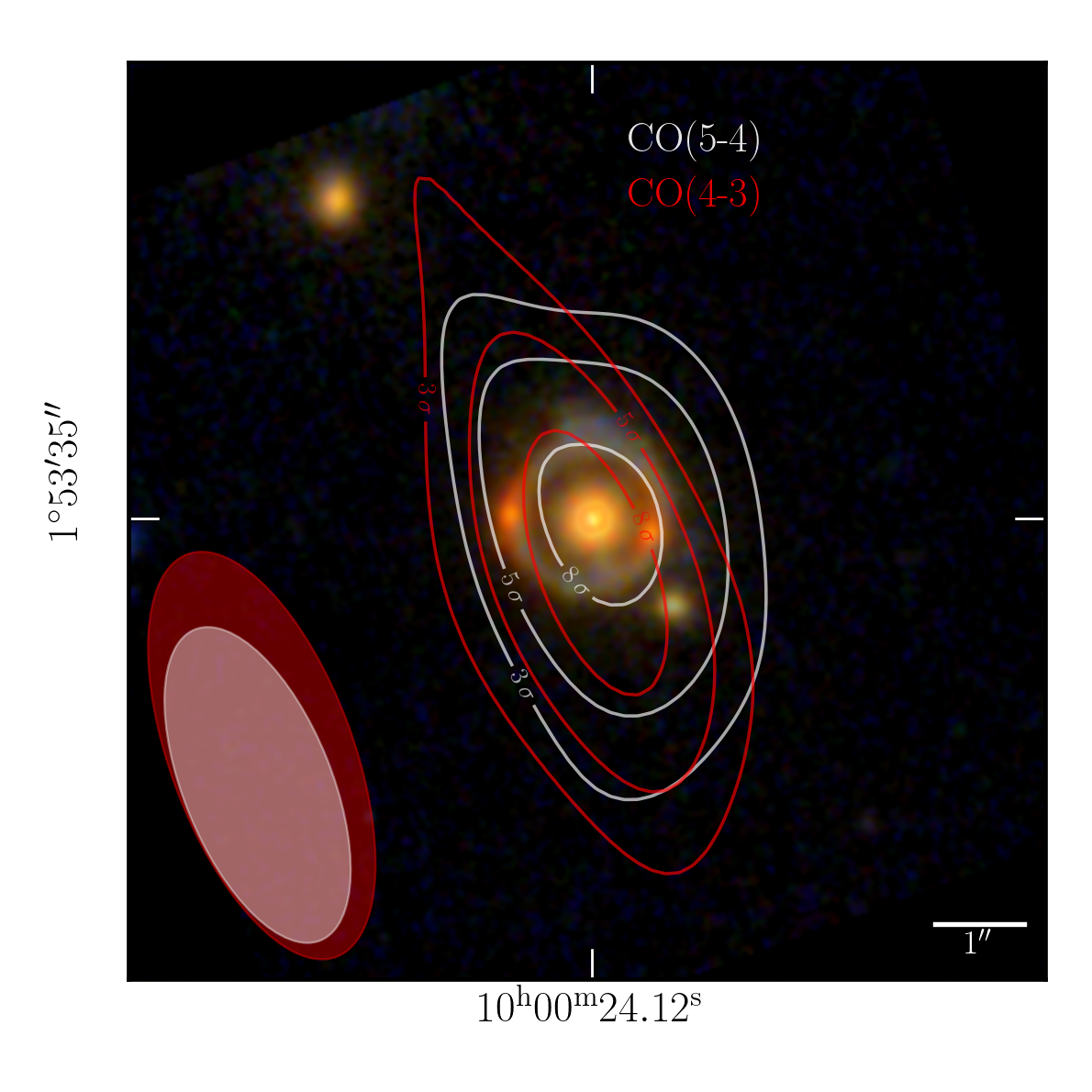}
    \vspace{5mm}
  \end{subfigure}
  \hspace{-6mm}
  \begin{subfigure}{0.45\textwidth}
    \includegraphics[width=\linewidth]{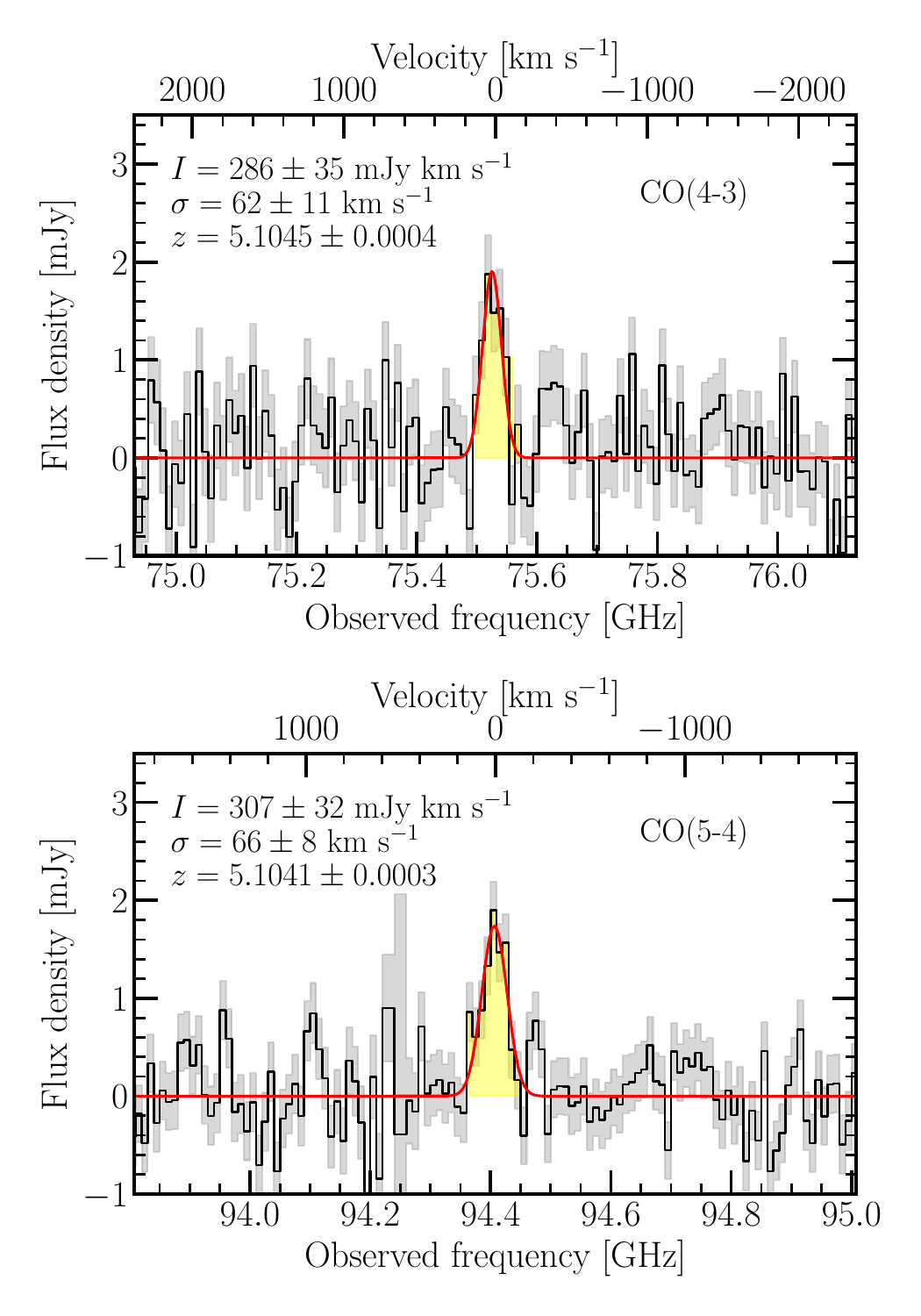}
  \end{subfigure}
  \caption{Observations of the COSMOS-Web ring from HST, JWST and NOEMA. Left panel shows a color image composed of F444W (red), F277W (green) and median stack of F150W+F814W (blue). Contours show the NOEMA CO maps at 3, 5, and $8\sigma$ levels. Red and white contours correspond to CO(4-3) and CO(5-4), respectively. Ellipses in the bottom-left corner indicate the corresponding beams. Right panels show the 1D spectra for CO(4-3) and CO(5-4) in 10MHz/channel resolution, and flux uncertainty is shown for each channel in gray. The lines are fitted with a Gaussian profile in red, the corresponding line intensity, velocity dispersion and redshift are presented in text.
  }
  \label{fig:map-and-spectra}
\end{figure*}

First, we conducted a MOS spectroscopic follow-up of the COSMOS-Web Ring using Keck/MOSFIRE in $H$ and $K$ bands in December 2023, as part of program N167 (PI: J. Kartaltepe). The target was observed for 2.19 hrs in $H$ and 1.72 hrs in $K$, with the slit oriented by 10° relative to the north so that it passed through the "blue" parts of the ring \citep[see][]{Mercier2024Ring} and the foreground lens. The objective was to observe [O\textsc{ii}]$\lambda\lambda$3726,3729 and Mg\,\textsc{ii}$\lambda\lambda$2796,2803 in the background source if it is at $z \approx 5$ \citep{Mercier2024Ring} or [O\textsc{ii}], [O\textsc{iii}]$\lambda\lambda$4959,5007, and H$\beta$ if at $z \approx 3$ \citep{vanDokkum2024Ring}.
The 2D data were reduced and the 1D spectra were extracted using two different reduction codes: 1) following the method described in \citet{Larson2022} and 2) using \textsc{PypeIt} \citep{pypeit:joss_pub,pypeit:zenodo}. These methods are meant to reduce the contamination between the foreground lens and the background ring. In particular, we extracted two 1D spectra of the background source: one for the northern part of the ring and the other one for the southern part. The two reductions yield the same results and we adopt the results from method 1) in this letter, since it is better tailored for high redshift targets.
We detect a $5.7\sigma$ emission line at $2.27\,\si{\micron}$ in $K$ band (Fig.~\ref{fig:mosfire}) in the northern spectrum with an equivalent width of $570\pm40$\,\AA. 
his line would correspond to [O\textsc{ii}] at $z = 5.10$ if $z\approx5$ \citep{Mercier2024Ring}, while it is not expected if $z\approx3$ \citep{vanDokkum2024Ring}.
However, this single line is not sufficient to robustly confirm the redshift given the photo-$z$ discrepancy and the low S/N (see a cautionary tale by \citealt{Jin2024lens}). To definitely pin down the redshift, we proposed observations with the IRAM Interferometer NOEMA (Northern Extended Millimeter Array).

The NOEMA observations were executed in December 2024 (ID: S24CE, PIs: M. Shuntov \& W. Mercier) with array configuration C and an on-source integration of 5.6 hrs. The frequency tunings are 72.7--82.8~GHz in the lower sideband (LSB) and 90.2--98.3~GHz in the upper sideband (USB), covering the CO(4-3), [CI](1-0) and CO(5-4) lines at $z=5.1$.
Data were reduced and calibrated using the GILDAS pipeline at the IRAM headquarters in Grenoble. We produce $uv$ visibility tables and perform analysis in the $uv$ plane, following the methods of \cite{ShuowenJin2019,Jin2022}. The resulting synthesized beams are $\sim 4.7''\times2''$ in LSB and $\sim 3.7''\times1.6''$ in USB, respectively.
The calibrated data reach a line sensitivity of $0.25$~mJy over a 100 km/s width at 80~GHz, and a continuum sensitivity of 6.8 $\mu$Jy. 
Two lines are detected with S/N$>8$, and dust continuum is not detected. Clean line maps are produced with the GILDAS/Mapping procedure \texttt{clark}.
The target is not spatially resolved in the NOEMA beams, we thus extracted the spectra by fitting a point source model in $uv$ space at the peak position of the line maps.

\section{Results} \label{sec:results}

\subsection{Redshift confirmation}

Fig.~\ref{fig:map-and-spectra} presents the NOEMA line detections. Two lines are detected at $\sim$75.5~GHz ($8.1\sigma$) and 94.4~GHz ($9.3\sigma$), which match explicitly to the frequencies of the CO(4-3) and CO(5-4) lines at $z=5.1$. By fitting a Gaussian profile to the two lines, we obtained consistent spectroscopic redshifts of $z_{\rm CO(5-4)}=5.1045\pm0.0004$ and $z_{\rm CO(4-3)}=5.1041\pm0.0003$, which the two CO lines have consistent line widths (${\rm FWHM} \sim150$~km~s$^{-1}$). We adopt a weighted average of the two CO redshifts, $z_{\rm spec}=5.1043\pm0.0004$. The redshift determined by CO lines is in excellent agreement with the redshift of MOSFIRE [O\textsc{ii}] (Fig.~\ref{fig:mosfire}). This therefore spectroscopically confirmed that the redshift of the lensed galaxy is $z_{\rm spec}=5.1043\pm0.0004$. This redshift validated the \textsc{eazy}photo-$z$ solution of $z_{\rm phot}=5.08_{-0.04}^{+0.06}$ by \cite{Mercier2024Ring}.
The \LePHARE\ and \Cigale\ $z_{\rm phot}\sim5.5$ and $z_{\rm phot}\sim5.3$ solutions in \cite{Mercier2024Ring} are slightly higher than the spec-$z$, but still agree within an uncertainty of $\Delta z/(1+z)<0.1$. 
Explicitly, our results disapprove the solution of $z_{\rm phot} = 2.98^{+0.42}_{-0.47}$ by \cite{vanDokkum2024Ring}.

We present the line intensity and velocity dispersion in Fig.~1 (\textit{right panels}). The dust continuum and [C\textsc{i}](1-0) line are not detected (Fig.~\ref{fig:spectra-full}), and we show the upper limits in Table \ref{tab:physpars}. 

\subsection{Physical properties}

\begin{figure}[t!]%
\setlength{\abovecaptionskip}{-0.2cm}
\setlength{\belowcaptionskip}{-0.2cm}
\centering
\includegraphics[width=1\columnwidth]{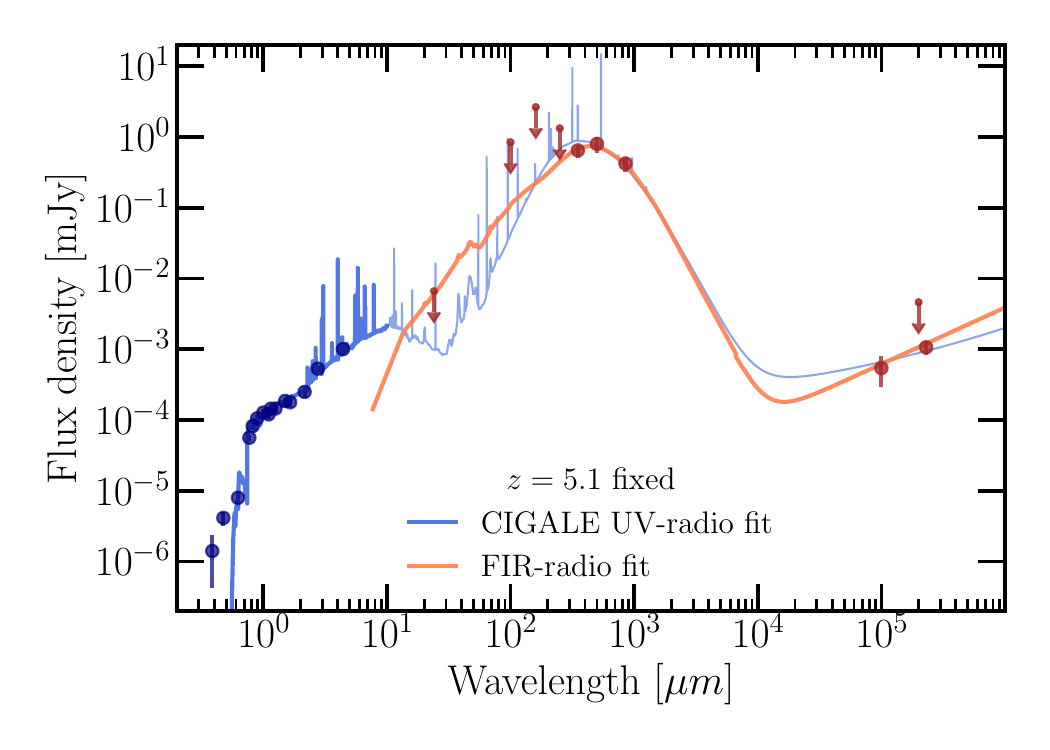}
\caption{Spectral energy distribution fitting of the background source. The UV-NIR and FIR-radio fits are carried out separately by fixing $z=5.1$. The best-fit models are shown as blue (\texttt{CIGALE}) and red (\texttt{MiChi2}) curves, respectively.
}
\label{fig:sed}
\end{figure}

With the confirmed redshift, we perform SED fitting to derive the physical properties of the lensed galaxy. We present the best-fit SEDs in Fig.~\ref{fig:sed}, where we fit the optical+NIR and FIR+radio data at fixed redshift $z=5.1$, with best-fit results listed in Table \ref{tab:physpars}.
We use the optical and NIR photometry obtained from the lens modeling using the \textsc{sl\_fit} code \citep{Gavazzi2008}, which measures the total and intrinsic flux of the background system, as presented by \citep{Mercier2024Ring}. This includes 15 broad bands from JWST/NIRCam, HST/ACS-F814W, CFHT/MegaCam-$u$, Subaru/HSC, and UltraVISTA.
We measured FIR and radio photometry by performing our Super-deblending pipeline \citep{Jin2018,Liu2018}. Because the foreground lens and the background source are not resolved in FIR and radio images, we use one prior to represent the whole system and fit the images with other sources in the same field. The deblended photometry includes MIPS 24~$\si{\micron}$, \textit{Herschel}, SCUBA2 850~$\si{\micron}$, VLA, and MeerKAT. We obtained S/N$>3$ detections from \textit{Herschel}/SPIRE 350~$\si{\micron}$ \& 500~$\si{\micron}$, SCUBA2 850~$\si{\micron}$ \citep{Simpson2019scuba2}, and MeerKAT 1.28~GHz \citep{Jarvis2016mightee,Hale2025MeerKAT}. Given that the foreground lens is quiescent \citep{Mercier2024Ring} and the lensed source shows red NIRCam colors with high dust attenuation ($E(B-V)\sim0.7$ on the two red knots), we assume that all FIR emissions come from the background source. 

We used two methods to fit the photometry: (1) we perform a panchromatic SED fitting spanning optical to radio wavelengths with \texttt{CIGALE} \citep{Boquien19} assuming an energy balance between UV and FIR; (2) we fit only the FIR+radio photometry using the \texttt{MiChi2} code \citep{Liu2021michi2} without including photometry shorter than 24~$\si{\micron}$.
As shown in Fig.~\ref{fig:sed}, the optical and NIR data are well fitted with \texttt{CIGALE} (\textit{blue curve}), which gives a ${\rm SFR=1172\pm209}~{\rm M_\odot}~{\rm yr}^{-1}$ (uncorrected for magnification). However, the FIR and radio fluxes are not well modeled, where the 350~$\si{\micron}$ flux was over-fit, and 850~$\si{\micron}$ and MeerKAT fluxes were under-fit. On the other hand, the \texttt{MiChi2} fits well to the FIR and radio photometry with a template assuming a typical IR-radio scaling \citep{Magnelli2015qIR} and a radio slope index of $-0.8$. This suggests that the radio emission also originates from the background source. The \texttt{MiChi2} fitting yields a best fit ${\rm SFR_{IR}=671\pm60~M_\odot ~yr^{-1}}$ uncorrected for magnification, while the dust mass and temperature are largely uncertain given the poor constraints on the dust SED peak and the Rayleigh-Jeans slope. We note that the high SFR from the \texttt{CIGALE} fitting could be due to the assumption of energy balance that is often not the case for DSFG SEDs. Therefore, we preferentially adopt the FIR+radio results from the \texttt{MiChi2} fitting and the optical+NIR results from the \texttt{CIGALE} fitting.

After correcting for the magnification $\mu=11.6$ \citep{Mercier2024Ring}, we found that the background source has a stellar mass of $M_{\star} \sim 1.8\times10^{10}~\si{\Msun}$ and an obscured ${\rm SFR_{IR}\sim 60~M_\odot ~yr^{-1}}$. The SFR is $\sim1.5$ times above the main sequence at $z=5$ constrained by recent JWST data \citep{Cole2025MS}. This suggests that the galaxy could be a starburst, however, it is still within the uncertainty of the main sequence.
Using [C\textsc{i}](1-0) as a gas tracer, we place a $2\sigma$ upper limit of the gas mass $M_{\rm gas}<5.1\times10^{9}\,\si{\Msun}$ using the [C\textsc{i}]-$M_{\rm gas}$ scaling from \cite{Valentino2018CI}. With the derived ${\rm SFR}_{\rm IR}=58\pm5\,\si{\Msun}$~yr$^{-1}$, we obtained a gas depletion time $\tau_{\rm dep}<100$~Myr, indicating a high star formation efficiency (SFE). This SFE is comparable to massive starburst galaxies at $z\sim5$ (e.g., \citealt{Ciesla20,Riechers2020z5,Brinch2025}), and 2--3 times higher than that of optically dark DSFGs at similar redshift (e.g., \citealt{Jin2022, Sillassen2025}). This again supports its starburst nature.
We note that the reconstructed source plane \citep{Mercier2024Ring} revealed a possible merger, in which a compact red core is likely to merge with two blue components. The high SFE is likely associated with the compact and dust-obscured component, which could be enhanced by a galaxy-galaxy merger.

\begin{table}[th!]
    \caption{Properties of the background source}
    \centering
    \setlength{\tabcolsep}{2pt}
    \renewcommand\arraystretch{1.5}
    \begin{tabular}{c c c c c c c c}
    \hline\hline
        RA  & $10^{\rm h}00^{\rm m}24.12^{\rm s}$\\
        Dec & $1^{\circ}53'35''$ \\
        $z_{\rm spec}$ & $5.1043\pm0.0004$\\
        $\mu$ & $11.6$\\
        $I_{\rm [CI](1-0)}$ [mJy~km~s$^{-1}$] &   $<5.7$ \\
        $S_{\rm 3.5mm}$ [$\mu$Jy] &   $<1.2$ \\
        $E(B-V)$ [mag] &  $0.33$ \\
        log$(M_\ast/M_\odot)$ & $10.26\pm0.15$ \\
        ${\rm SFR}_{\rm IR}$ [${\rm M_\odot\,yr^{-1}}$] & $58\pm5$\\
        $M_{\rm gas,[CI]}$ [${\rm 10^{9} M_\odot}$] & $<5.1$\\
        ${\rm SFE}$ [${\rm Gyr^{-1}}$] & $>10.2$\\
        ${\tau_{\rm dep}}$ [Myr] & $<96$\\
        \hline\hline
    \end{tabular}
    {\\Notes: All parameters have been corrected for magnification $\mu$. Upper limits are $2\sigma$.}
    \label{tab:physpars}
\vspace{-0.3cm}
\end{table}

\section{Discussion \& Conclusion} \label{sec:Discussion-Conclusions}

Using NOEMA and Keck/MOSFIRE spectroscopy, we have unambiguously confirmed the background source of the COSMOS-Web ring to be at $z=5.104$. 
Remarkably, this is the most distant Einstein ring system known to date, in both the foreground lens ($z\sim2$) and the background source ($z=5.1$), thanks to the exquisite capability of JWST to detect distant galaxy-galaxy lenses (Nightingale et al. in prep., Mahler et al in prep., Hogg et al. in prep.).
Using multi-wavelength photometry spanning NIR to radio bands, we revealed that the background galaxy is a dust-obscured starburst (${M_{\star} \sim 1.8\times10^{10}\,\si{\Msun}}$, ${\rm SFR_{IR}\sim 60\,\si{\Msun} ~yr^{-1}}$) with high star-formation efficiency ($\tau_{\rm dep}<100~$Myr).

The redshift confirmation of the background source at $z=5.10$ also supports the conclusion of \cite{Mercier2024Ring} that dark matter and stellar mass fully explain the lens mass budget: lensing estimates with a $z\sim5$ background source yield a total lens mass within the Einstein radius $M_{\rm tot}(<\theta_{\rm Ein})\sim3.8 \times 10^{11} \, \si{\Msun}$, out of which $M_{\star}(<\theta_{\rm Ein}) \sim1.2 \times 10^{11} \, \si{\Msun}$ come from stars, and $M_{\rm dm}(<\theta_{\rm Ein}) \sim 2.6 \times 10^{11} \, \si{\Msun}$ come from dark matter. The results are consistent with stellar-to-halo mass scaling relation \citep{Shuntov2022}, indicating that stars and dark matter fully account for the total mass budget of the lens, without the need of modifying IMF or the dark matter profile \citep[as suggested by][based on a $z_{\rm phot}\sim3$ solution for the background source]{vanDokkum2024Ring}.

The upcoming JWST NIRSpec/IFU observations (ID: 5883, PI: R. Gavazzi) will confirm the lens redshift and allow for detailed dark-matter substructure reconstruction of the lens. Meanwhile, the IFU data will detect rest-frame optical lines of the background source, enabling measurements of metallicity, dust attenuation, and potential AGN signatures. 
Unfortunately, no ALMA observations yet exist for this unique system. Future ALMA follow-up (e.g., CO, [CII], continuum) would be essential to spatially resolve the background source and reveal detailed dust/gas/stellar geometry, ISM conditions, and spatially resolved kinematics.

\begin{acknowledgements}

This work is based on observations carried out under projects number S24CE with the IRAM Interferometer NOEMA. IRAM is supported by INSU/CNRS (France), MPG (Germany) and IGN (Spain). We are grateful for the help received from IRAM staﬀ during observations and data reduction.
This work was supported by a NASA Keck PI Data Award, administered by the NASA Exoplanet Science Institute. Data presented herein were obtained at the W. M. Keck Observatory from telescope time allocated to the National Aeronautics and Space Administration through the agency's scientific partnership with the California Institute of Technology and the University of California. The Observatory was made possible by the generous financial support of the W. M. Keck Foundation. The authors wish to recognize and acknowledge the very significant cultural role and reverence that the summit of Maunakea has always had within the Native Hawaiian community. We are most fortunate to have the opportunity to conduct observations from this mountain.
The Cosmic Dawn Center (DAWN) is funded by the Danish National Research Foundation (DNRF140).
SJ acknowledges the support from the European Union’s Horizon Europe research and innovation program under the Marie Sk\l{}odowska-Curie Action grant No. 101060888. 
SJ and GEM acknowledges the Villum Fonden research grants 37440 and 13160.
JSK and AAK acknowledge support from the National Science Foundation under Grant No. 2009572 and from NASA under award No. 80NSSC22K0854. 
    This work was made possible thanks to the CANDIDE cluster at the Institut d’Astrophysique de Paris, which was funded through grants from the PNCG, CNES, DIM-ACAV, and the Cosmic Dawn Center; CANDIDE is maintained by S. Rouberol. 
    The French contingent of the COSMOS team is partly supported by the Centre National d’Etudes Spatiales (CNES). OI acknowledges the funding of the French Agence Nationale de la Recherche for the project iMAGE (grant ANR-22-CE31-0007). 

\end{acknowledgements}

%
%

\bibliographystyle{aa_url}
\bibliography{biblio.bib}

\begin{appendix}
\onecolumn

\section{Supporting material}


\begin{figure*}[ht!]
\centering
\includegraphics[width=0.9\textwidth]{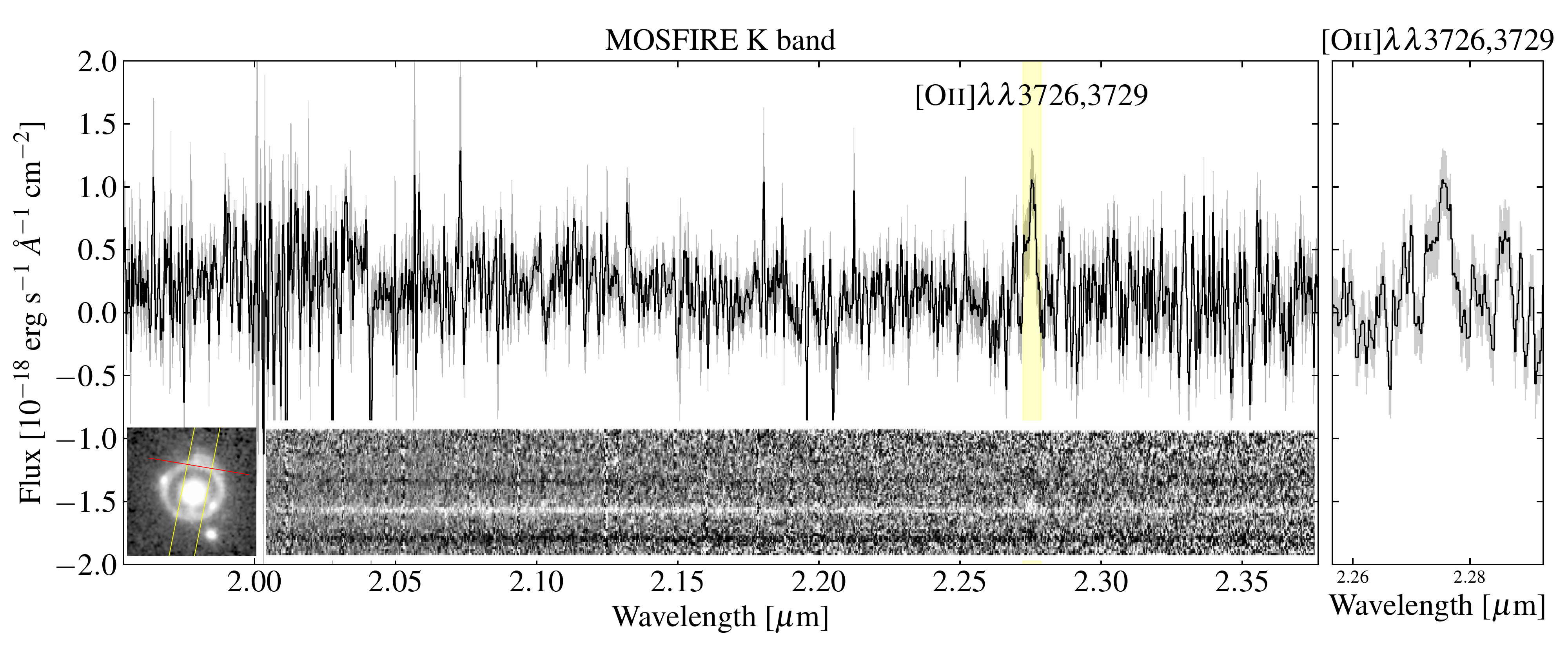}
\caption{MOSFIRE spectrum of the Einstein ring. 
A zoom-in on the [OII] doublet (S/N$=5.7$) is shown on the right. The inset shows the slit configuration (yellow box) the location where the spectrum is extracted (red line) and the 2D spectrum. 
}
\label{fig:mosfire}
\end{figure*}


\begin{figure*}[ht!]
\centering
\includegraphics[width=0.99\textwidth]{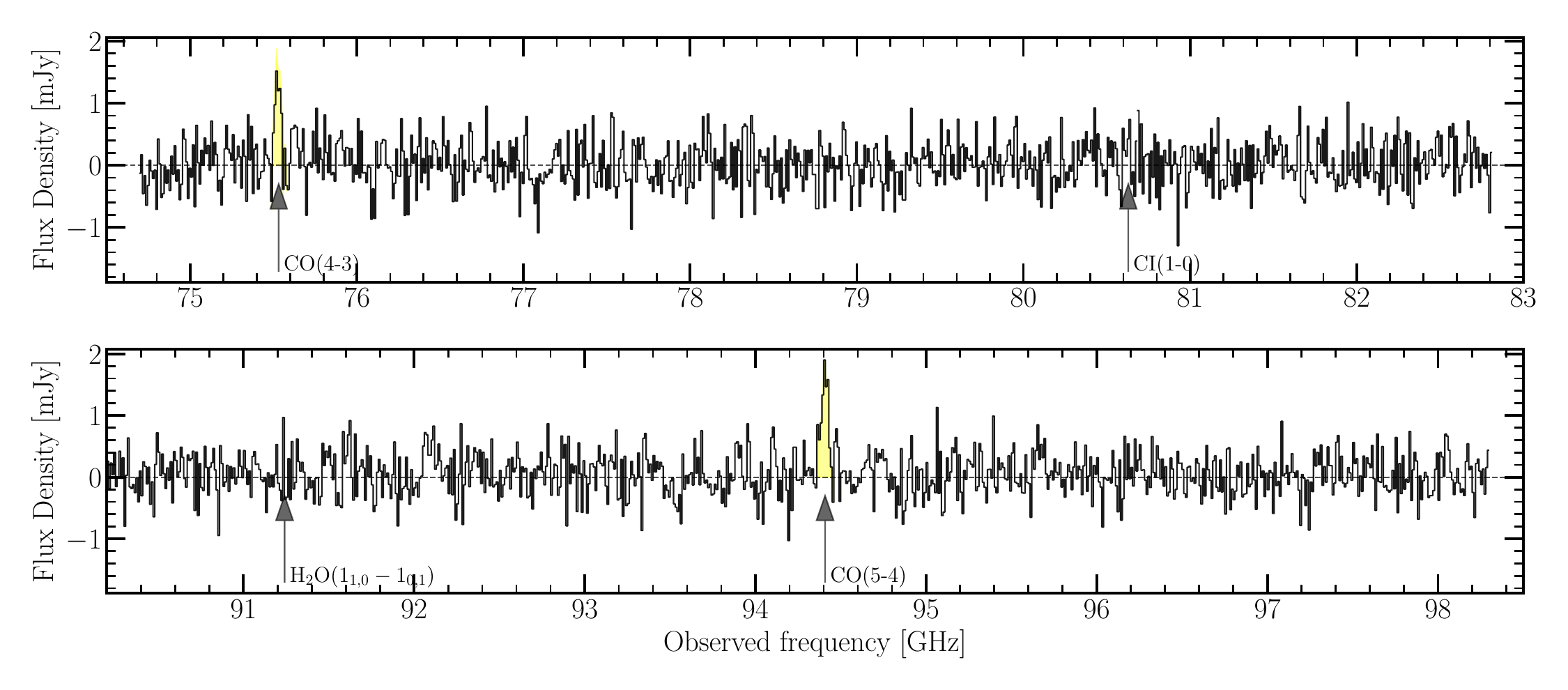}
\caption{Full NOEMA spectra in the LSB (top) and USB (bottom). The arrows mark the CO and [CI] frequencies at $z=5.1$. We highlight the CO(4-3) and CO(5-4) lines in yellow, while [CI](1-0) and H$_2$O($1_{1,0}-1_{0,1}$) show no detection.}
\label{fig:spectra-full}
\end{figure*}



\end{appendix}

\end{document}